\documentclass[manuscript,screen]{acmart}
\AtBeginDocument{%
  \providecommand\BibTeX{{%
    \normalfont B\kern-0.5em{\scshape i\kern-0.25em b}\kern-0.8em\TeX}}}

\copyrightyear{2023}
\acmYear{2023}
\setcopyright{acmlicensed}
\acmConference[RecSys '23]{Seventeenth ACM
Conference on Recommender Systems}{September 18--22, 2023}{Singapore,
Singapore}
\acmBooktitle{Seventeenth ACM Conference on Recommender Systems (RecSys
'23), September 18--22, 2023, Singapore, Singapore}
\acmPrice{15.00}
\acmDOI{10.1145/3604915.3608848}
\acmISBN{979-8-4007-0241-9/23/09}

\begin{document}


\title[Linking Interactions and User's Propensity Towards Multi-Objective Recommendations]{Looks Can Be Deceiving: Linking User-Item Interactions and \\User's Propensity Towards Multi-Objective Recommendations}

\author{Patrik Dokoupil}
\email{patrik.dokoupil@matfyz.cuni.cz}
\orcid{0000-0002-1423-628X}
\affiliation{%
  \institution{Faculty of Mathematics and Physics, Charles University, Prague}
  \country{Czechia}
}
  
\author{Ladislav Peska}
\email{ladislav.peska@matfyz.cuni.cz}
\orcid{0000-0001-8082-4509}
\affiliation{%
  \institution{Faculty of Mathematics and Physics, Charles University, Prague}
  \country{Czechia}
}

\author{Ludovico Boratto}
\email{ludovico.boratto@acm.org}
\orcid{0000-0002-6053-3015}
\affiliation{%
  \institution{University of Cagliari}
  \country{Italy}
  }
\renewcommand{\shortauthors}{Dokoupil, Peska and Boratto}

\begin{abstract}

Multi-objective recommender systems (MORS) provide suggestions to users according to multiple (and possibly conflicting) goals. When a system optimizes its results at the individual-user level, it tailors them on a user's propensity towards the different objectives. Hence, the capability to understand users' fine-grained needs towards each goal is crucial. In this paper, we present the results of a user study in which we monitored the way users interacted with recommended items, as well as their self-proclaimed propensities towards relevance, novelty, and diversity objectives. The study was divided into several sessions, where users evaluated recommendation lists originating from a relevance-only single-objective baseline as well as MORS. We show that, despite MORS-based recommendations attracting fewer selections, their presence in the early sessions are crucial for users' satisfaction in the later stages. Surprisingly, the self-proclaimed willingness of users to interact with novel and diverse items is not always reflected in the recommendations they accept. 
Post-study questionnaires provide insights on how to deal with this matter, suggesting that MORS-based results should be accompanied by elements that allow users to understand the recommendations, so as to facilitate the choice of whether a recommendation should be accepted or not. Detailed study results are available at \url{https://bit.ly/looks-can-be-deceiving-repo}.



\end{abstract}





\maketitle

\section{Introduction}
{\bf Motivation and context. }{\em Beyond-accuracy objectives} are gaining more and more attention in Recommender Systems (RSs). Indeed, it is now paramount to pair recommendation effectiveness with properties that account for user perspectives (such as novelty and diversity~\cite{CastellsHV22} or consumer fairness~\cite{BorattoM21}), or that are aligned with the recommended items (such as behavioral biases or provider fairness~\cite{BorattoM21}).
{\em Multi-objective recommender systems} (MORS) support this paradigm by generating results that account for multiple properties~\cite{ZhengW22}. Recent literature has studied how to account for multi-objective goals from different angles. The user perspective was tackled by \citet{DBLP:conf/www/LiCFGZ21}, which balance recommendation accuracy for users with different levels of activities. From an item perspective, \citet{DBLP:conf/wsdm/GeZYPHHZ22} proposed an approach to balance item relevance and exposure. Considering both the user and item perspectives, \citet{DBLP:conf/sigir/NaghiaeiRD22} propose a re-ranking approach to account for  consumer and provider fairness. Other studies blend the multiple objectives into a single function, in order to obtain a Pareto-optimal solution~\cite{DBLP:conf/recsys/LinCPSXSZOJ19, wu2022multifr}. Recent advances have also proposed MORSs in sequential settings, by optimizing the results for accuracy, diversity, and novelty~\cite{DBLP:conf/wsdm/StamenkovicKAXK22}.

MORS can account for  multiple objectives at the {\em aggregate} level, by balancing these objectives over the entire user base (e.g., the system is capable of offering a certain level of diversity), or at the {\em individual} level, by matching the beyond-accuracy needs of each user in a different way (e.g., the recommendations of one user might intentionally be more diverse than those of another one)~\cite{Jannach22}.
MORS that operate at the individual level have optimized the recommendation process mainly via online interactions, such as conversational approaches \cite{10.5555/3327546.3327641,GAO2021100} or via critiquing \cite{WANG2020106369,elahiinteractive}, but approaches aiming at learning individual propensities from past interactions also exist, e.g., \cite{JUGOVAC2017321}.


\vspace{1mm}
\noindent {\bf Open issues. } Even though MORS that operate at the individual level have as a goal the optimization of the needs of each user, their functioning and evaluation either requires continuous interaction with the users or is based on offline data without any feedback from the users. Having these two extremes as the only options leads to two main questions that so far remain unanswered. At the RS functioning level, we need to understand how to incorporate the propensity of users towards certain beyond-accuracy properties into the recommendation process. This is not possible in offline approaches, while online ones work until a recommendation is accepted (i.e., the conversation or the critiques stop appearing). At the evaluation level, 
we do not know to what extent the recommendations accepted by the users are driven by these beyond-accuracy goals. Hence, {\em understanding directly from the users their propensity towards beyond-accuracy goals and how they should be reflected in the recommendations} is a key open problem for the functioning of MORS that operate at the individual level.

\vspace{1mm}
\noindent {\bf Our contributions. } To address the aforementioned issues, we present the results of a user study aimed at linking the self-proclaimed propensity of users towards relevance, novelty, and diversity criteria with their actual acceptance of provided recommendations. 
In particular, we asked users to iterate through several recommending sessions in the Movie domain. We confronted them with results of a relevance-based single-objective RS and two MORS variants balancing relevance, novelty, and diversity criteria. We further allowed them to tune MORS by defining their propensity towards the aforementioned criteria.
Therefore, for the first time in the literature, we can link (i) the propensity of the users to interact with items characterized by certain beyond-accuracy properties, with (ii) their propensity to accept recommendations offering these same properties.

Our results provide interesting insights into how users' propensity towards beyond-accuracy goals can be reflected in the individual-level MORS.
Indeed, despite the users' self-declared propensity towards multi-objective goals, single-objective recommendations attracted more selections than those generated by MORS. In the evaluation, we argue that the presence of MORS recommendations (and selections) is crucial for long-term user satisfaction.
We also discovered that users' selection behavior exhibits interesting deviations from the distributions induced by displayed items (impressions) and propensities towards individual objectives (weights). 
Indeed, in the case of single-objective RS, users on average selected items with lower estimated relevance scores than the average of recommended items. 
Likewise, in the case of MORS, users selected less diverse and novel items with higher estimated relevance than what was the average of those metrics w.r.t. the recommended items. This propensity towards relevant items in MORS-based recommendations happened regardless of the fact that the users could manually fine-tune the level of novelty and diversity. We briefly analyze the possible causes of these phenomena and suggest plausible mitigation strategies.
\section{User Study Design}\label{sec:study}
The study was conducted online\footnote{\url{https://bit.ly/looks-can-be-deceiving-study}} and consisted of the following steps: informed consent and basic demographics, preference elicitation, recommendation sessions (8x), and a post-study questionnaire.

\noindent {\bf Dataset and pre-processing. }The study was conducted on top of the  MovieLens-Latest dataset~\cite{10.1145/2827872}, which was selected for its relative novelty and high familiarity with the movie domain. The dataset was utilized in two ways: to populate collaborative filtering algorithms and as a starting point to gather item metadata. In order to comply with the gathered user selections, the feedback was binarized. Furthermore, to only focus on the relevant portion of the dataset, we filtered out movies released before 1990, ratings older than 2010, movies that have less than 50 ratings per year, users with less than 100 ratings, and movies without ratings. This resulted in 9K users, 2K movies, and 1.5M ratings. In order to properly visualize the items, additional metadata were collected from respective IMDb profiles: movie descriptions, posters, and links to movie trailers.

\noindent {\bf Recommender systems. }
In the study, three RSs variants were evaluated (one single-objective and two multi-objective), denoted as \textit{Beta}, \textit{Gamma}, and \textit{Delta}. \textit{Beta} (single-objective baseline) follows a generalized matrix factorization ~\cite{4781121} example from tf.recommenders\footnote{\url{https://www.tensorflow.org/recommenders/examples/basic_retrieval}}. We used the embedding size of 32 and 5 training epochs. \textit{Gamma} and \textit{Delta} utilized the predictions of \textit{Beta} as its relevance component, but additionally incorporated also diversity and novelty viewpoints. In particular, \textit{Delta} utilized RLProp algorithm \cite{10.1145/3477495.3531787} and \textit{Gamma} utilized incremental weighted average \cite{10.1145/3477495.3531787}.\footnote{I.e., recommended items were selected one by one, while their marginal gains were iteratively updated.} Both algorithms were parameterized by the user's propensity towards individual objectives (described in Sec. \ref{sec:importance_weights}).

\textit{Beta} algorithm was first trained on the MovieLens dataset and then fine-tuned separately (i.e., each study participant received their own private copy of the algorithm). Fine-tuning was done after the preference elicitation step as well as after each recommending session. Note that since the \textit{Beta} algorithm was utilized as a source in both \textit{Gamma} and \textit{Delta}, the feedback received on all recommended items was utilized for \textit{Beta} fine-tuning. Also note that to enhance engagement and coverage, we prohibited repeated recommendations of items that were previously shown to the user.

\subsection{Study flow}

In the \textbf{\textit{initial}} phase, users received a description of the study and were asked for basic demographics (e.g., gender, age, education) as well as to provide informed consent on the study procedure and publication of anonymized results.

In the \textbf{\textit{preference elicitation}} phase, participants were asked to select previously known and liked movies out of a randomized list. Depicted movies were sampled on the basis of three objective criteria: overall relevance, novelty, and diversity. For each criterion, we constructed bins of movies with high and low values, and from each bin, we randomly sampled four movies.\footnote{The overall relevance was considered w.r.t. average user profile. The novelty was defined as the item's mean popularity complement. The diversity was defined as collaborative intra-list diversity (ILD) w.r.t. the already selected movies from popularity and novelty bins.} This procedure aimed to minimize the historical biases present in the source data. Note that users were allowed to load more movies (based on the same procedure) as well as search for a specific movie manually. There were no strict limits on the volume of selected movies, but participants were instructed to try to select at least 5-10 movies (the median volume of actually selected items at this phase was 10).

During each of eight \textit{\textbf{recommendation sessions}}, the results of two RS were shown to the user. Each time we depicted an output of the single-objective RS (\textit{Beta}) accompanied by one of the MORS (either \textit{Gamma} or \textit{Delta}). Recommendation lists were kept separated, and displayed at randomized positions. The procedure for choosing the MORS variant was as follows. Before the first session, either \textit{Delta} or \textit{Gamma} RS was selected at random. This algorithm is then used in the first four sessions, while in the last four sessions, we switch to the other MORS variant. As such, algorithm-specific sequence-aware patterns can be observed and the usage MORS variant can be considered as \textit{within-subject} variable.\footnote{Merely the ordering of MORS variants is a \textit{between-subject} variable. This was a compromise solution. During study dry-runs, we observed that showing all three recommendation lists at once imposed an excessive cognitive burden on the users. On the other hand, making MORS variant a between-subject variable could introduce an excessive user-specific variance in the results.}

At each recommending session, we asked study participants to provide both implicit feedback (i.e., select items that they would consider watching tonight) and to provide explicit feedback (i.e., rate the overall performance of depicted RSs on a one-to-five stars scale). After completing the feedback phase, participants were also allowed to modify their propensity (i.e., weights) towards individual objective criteria. This was conducted via a slider depicting the current values for each objective and forcing it to maintain a unit sum of all objectives. 

Finally, in the \textit{\textbf{post-study questionnaire}}, we asked participants to fill in responses (on a 5-point Likert scale) to a series of questions regarding both the general performance of RS as well as questions specifically targeting the user interface (UI) for changing objective weights. Questions were inspired by the ResQue framework \cite{10.1145/2043932.2043962}, but extended to also cover the specifics of the UI for criteria propensity setting (see \url{https://bit.ly/looks-can-be-deceiving-repo} for details). The questionnaire also contained several attention checks to remove unreliable participants.

\subsection{Considered objectives and their importance weights}
\label{sec:importance_weights}
Both the \textit{Gamma} and \textit{Delta} RSs aim to incrementally construct the list of recommendations w.r.t. several objective criteria. In particular, they utilize the normalized marginal gains (NMG) individual items provide in terms of these objectives. In this paper, we focused on relevance, novelty, and diversity, defined as follows. For relevance, we considered the sum of estimated relevance scores (predicted by \textit{Beta} algorithm) as an objective, so the marginal gain of each item was its own relevance score: $MG_{i,rel} = \hat{r}_{u,i}$. Normalization is then applied as empirical cumulative distribution function (CDF) w.r.t. all items' marginal gains (see \cite{10.1145/3477495.3531787} for more details). Similarly, marginal gain w.r.t. novelty was defined as item's mean popularity complement \cite{epc}: $MG_{i,nov} = 1 - |u \in U: r_{u,i}\text{ exists}| / |U|$, where $r_{u,i}$ is the feedback of user $u$ on item $i$ and $U$ is the set of all users. Marginal gain w.r.t. diversity is defined as the mean collaborative distance of the item to the list of already selected recommendations:\footnote{I.e., the diversity objective corresponds to the incremental collaborative intra-list diversity, ILD \cite{bradley2001improving}.} $MG_{i,div} = \frac{1}{|L|}\sum_{\forall j \in L} d(i,j)$, where $L$ is a list of already selected recommendations and $d(i,j)$ is a distance metric -- cosine distance on items' ratings in our case. 

Both the \textit{Gamma} and \textit{Delta} algorithms used the propensity weights assigned to individual objectives. These were iteratively modified by the users after each session, but their initial values had to be trained based on the data from preference elicitation. We used a similar procedure to \cite{JUGOVAC2017321}. In particular, we calculated the normalized marginal gains (NMG) for each objective and each selected movie. Note that because the user's profile was not established yet, relevance gain was calculated as the mean estimated relevance of the selected items w.r.t. all train set users. Diversity gain was calculated as the mean distance of the selected items from all the displayed ones, and novelty gain remains unchanged. Gains of all selected movies were normalized via CDF defined on the population of all displayed movies. Final estimated propensities were obtained as the mean of all items' NMGs and linearly scaled to unit sum.

\section{Study Results}\label{sec:results}
The study was conducted in April 2023. In total, 120 participants were recruited using the \url{Prolific.co} service. Participants were pre-screened for fluent English, no less than 10 previous submissions, and 99\% approval rate. Twelve users did not finish the study and, in addition, we rejected 2 participants due to failed attention checks, which resulted in 106 completed participations.

\begin{table*}[tb]
\caption{Overall results w.r.t. user feedback and normalized marginal gains of recommended and selected items. Best results are in bold, while significantly inferior results (T-test p-value < 0.05) are denoted with an asterisk (*).}
\label{tab:overall_results}
\begin{tabular}{r|rr|rrr|rrr}
\toprule
          & \multicolumn{2}{c}{Feedback}  & \multicolumn{3}{c}{Impressions} & \multicolumn{3}{c}{Selections}  \\
Algorithm & Selections ratio & Mean rating & $NMG_{rel}$ & $NMG_{div}$ & $NMG_{nov}$ & $NMG_{rel}$ & $NMG_{div}$ & $NMG_{nov}$ \\
\midrule
Beta      &            \textbf{0.37}     &           \textbf{3.20}  &     \textbf{0.980} &          *0.251 &        *0.653   &   \textbf{0.974} &          *0.246 &        *0.635 \\
Gamma     &            *0.19     &           *2.34  &     *0.875 &          \textbf{0.787} &        \textbf{0.849}   &   *0.910 &          \textbf{0.689} &        \textbf{0.816} \\
Delta     &            *0.26     &           *2.68  &     *0.913 &          *0.659 &        *0.790   &   *0.943 &          *0.515 &        *0.719 \\    
\bottomrule
\end{tabular}
\end{table*}

Table \ref{tab:overall_results} depicts the overall study results. 
It can be seen that \textit{Beta} (relevance-only baseline) significantly outperformed both MORS variants w.r.t. total volume of selections as well as mean algorithm ratings. Out of the two MORS variants, \textit{Delta} obtained significantly more selections (Fisher's exact test p-value: 2.6e-15) as well as significantly higher average ratings than \textit{Gamma} (T-test p-value: 8.6e-6). Note that both implicit and explicit feedback modalities were correlated, but there were some discrepancies (Pearson's correlation: 0.63). We also checked several other statistics with rather expectable results: The volume of selections slightly drops for subsequent sessions (up to 28\% drop), top-ranked items were selected more often than lower-ranked (up to 33\% drop), etc. Some additional details are available from \url{https://bit.ly/looks-can-be-deceiving-repo}.
Based on these initial results, we formulated the following questions: 
\begin{itemize}
    \item \textbf{RQ1.} Are there some qualities, in which evaluated MORS  improve over the single-objective baseline? If so, what is the long-term impact of these qualities on user satisfaction?
    \item \textbf{RQ2.} What are the possible causes of the inferior performance of MORS? Could this be somehow mitigated?
\end{itemize}

\subsection{Beyond-accuracy 
objectives and their long-term impact}
\label{sec:results_long_term}
In order to answer RQ1, we focused on the beyond-accuracy criteria of recommended (impressions), but also selected items. A natural choice to start with are the normalized marginal gains of considered objectives. By inspecting Table \ref{tab:overall_results}, one can observe that both \textit{Gamma} and \textit{Delta} clearly outperformed \textit{Beta} in terms of $NMG_{nov}$ and $NMG_{div}$ for impressions as well as selections. The increased impression-level novelty and diversity is a direct consequence of recommendations construction, while the selection-level increase indicates that users (to some extent) followed the distribution of recommended items. Obtained results also corresponded to other novelty and diversity metrics: collaborative ILD (0.876 for \textit{Beta} vs. 0.988 for \textit{Gamma} vs. 0.961 for \textit{Delta}), content-based (genre-based) ILD (0.323 vs. 0.385 vs. 0.369), genre coverage (0.484 vs. 0.482 vs. 0.496), mean popularity complement (0.973 vs. 0.996 vs. 0.989), temporal novelty (0.647 vs. 0.932 vs. 0.856). Significant differences were also obtained on selections: collaborative ILD (0.868 vs. 0.964 vs. 0.946),  mean popularity complement (0.965 vs. 0.992 vs. 0.990), and temporal novelty (0.625 vs. 0.896 vs. 0.859).

However, although the above-mentioned results are interesting, it is yet to be shown whether they have a practical impact. To do so, we focused on (i) the long-term user satisfaction as a function of users' acceptance of MORS recommendations in early sessions, and (ii) the impact of MORS-based selections on training single-objective RS. 

\vspace{1mm}
\noindent {\bf Impact of MORS acceptance on long-term user satisfaction.}
Let us (optimistically) assume that eight recommendation sessions constitute a sufficient base for a long-term evaluation. Our analysis is based on dividing the sessions into early (i.e., \textit{head}) and late (i.e., \textit{tail}). Then, we measure whether the adoption of MORS recommendations in the \textit{head} had a measurable impact on user satisfaction in the \textit{tail}.

In particular, we considered the size of the \textit{head} to be one to four first sessions and 
defined three metrics (w.r.t. \textit{head}) to describe a user's MORS adoption: the volume of selections on single-objective RS, the volume of selections on multi-objective RS, and the ratio of multi-objective selection on all selections (in the results; we denote them as \textit{\#SORS}, \textit{\#MORS}, \textit{MORS ratio} respectively). For each metric, we divided users into two groups: users with above-median values and with values below-or-equal to the median (denoted as \textit{High} and \textit{Low} clusters). In order to get finer-grained results, we also applied a pre-processing of users to separately consider those who had high or low total volumes of selections in the head segment (denoted as \textit{high-selections}, \textit{low-selections}, and \textit{all users} for no pre-processing). In the \textit{tail} section, we considered the total volume of selections (\textit{\#selections}) and the \textit{mean rating} of the provided recommendation lists.

\begin{table*}[tb]
\caption{Results of the long-term impact of MORS selections (i.e., \textit{tail} sessions). For the sake of space, we only depict the \textit{head} section sizes of two and three. Significantly better results (one-sided T-test p-value < 0.05) are bold, while an asterisk (*) denotes p-values < 0.01.}
\label{tab:head_tail}
\begin{tabular}{l|rr|rr|rr||rr|rr|rr}
\toprule
                & \multicolumn{6}{c}{Head size: 2}                                                         & \multicolumn{6}{c}{Head size: 3} \\
    & \multicolumn{2}{c}{\#MORS} & \multicolumn{2}{c}{\#SORS} & \multicolumn{2}{c}{MORS ratio} & \multicolumn{2}{c}{\#MORS} & \multicolumn{2}{c}{\#SORS} & \multicolumn{2}{c}{MORS ratio} \\
\textbf{Mean ratings}  & Low         & High         & Low         & High         & Low           & High           & Low         & High         & Low         & High         & Low           & High           \\
\midrule
Low-selections  &  2.56 &  2.70 & 2.59 &  2.66 &  2.53 &  2.73  &   2.54 &  2.61 & 2.56 &  2.59 & 2.32 &  *\textbf{2.83} \\
High-selections &  2.78 &  \textbf{3.07}  & 2.89 &  2.95 &  2.72 &  *\textbf{3.12}  &  2.86 &  \textbf{3.15} & 2.99 &  2.96 &   2.79 &  \textbf{3.17} \\
All users           &  2.61 &  *\textbf{2.97}&  2.68 &  2.86 & 2.61 &  *\textbf{2.92}  & 2.50 &  *\textbf{3.02} &2.62 &  \textbf{2.91}  &  2.57 &  *\textbf{2.94} \\     
\midrule
\textbf{\# selections}  & Low         & High         & Low         & High         & Low           & High           & Low         & High         & Low         & High         & Low           & High           \\
\midrule
Low-selections  & 21.7 &  \textbf{31.6} & 19.8 &  *\textbf{33.5} & 28.3 &  23.5  &   18.5 &  24.4& 16.7 &  *\textbf{26.1}& 22.6 &  19.9 \\
High-selections &  37.1 &  \textbf{47.8} &  39.1 &  45.8 &  37.0 &  *\textbf{47.9}&   33.7 &  \textbf{40.5} & 33.2 &  \textbf{40.2}  &  34.5 &  38.2 \\
All users           &   28.1 &  *\textbf{41.7}&  27.3 &  *\textbf{40.9} & 33.7 &  34.1  &  21.2 &  *\textbf{35.2}  &  22.1 &  *\textbf{34.8} & 27.4 &  28.7 \\ 
\bottomrule
\end{tabular}
\end{table*}

Table \ref{tab:head_tail} shows the results of the long-term impact evaluation. The main outcomes are as follows. For \textit{\#MORS}, the \textit{High} cluster exhibited better values of both \#selections and mean ratings. However, the inherent flaw is that the same trend appeared in the \textit{head} section as well\footnote{I.e., users who made more MORS selections also made more selections and provided higher ratings in general.}. This tendency is maintained throughout the study, seemingly without major fluctuations. Therefore, we assume that the high \textit{\#MORS} cluster (w.r.t. \textit{head}) merely identifies a cluster of more overall engaged users.

In contrast, \textit{High} and \textit{Low} clusters w.r.t. \textit{MORS ratio} exhibited much more similar performance in the \textit{head} section (at least for shorter \textit{heads} -- see further). In the \textit{tail} sections, users of the \textit{High} cluster provided on average significantly higher ratings than the users of the \textit{Low} cluster. To our surprise, the impact on the volume of the selections was much smaller, often insignificant, or even negative. That is, despite selecting similar (or even lower) volumes of items, users of the \textit{High} cluster were in general more satisfied with provided recommendations. Note that the impact is incremental and rather fast. While for the \textit{head} sizes of one and two, there is no substantial difference in the performance of both clusters w.r.t. head, this gradually changes and already for the head size of four, this became noticeable. 

Interesting results were also obtained for \textit{\#SORS}. While the \textit{high} cluster almost always exhibited a higher volume of all selections in the \textit{tail}, the improvements w.r.t. mean ratings were smaller, mostly insignificant, or even negative. This was despite the fact that quite often, \textit{high} user clusters were associated with higher mean ratings in the head section. We read these results in such a way that, despite being satisfied in the early stages, users who mostly adopted single-objective recommendations struggle to find sufficiently interesting/satisfying recommendations in the later stages. This is despite the fact that there are many ``somewhat relevant'' items (thus the higher volume of selections). As for the pre-processing variants, the general trend was similar for both sub-groups, but the differences were more pronounced on \textit{high-selection} cluster of users. We assume that this is a natural consequence of the fact that more data is being supplied for fine-tuning.

Overall, we may conclude that early adoption of MORS-based recommendations led to higher satisfaction later on. This is further corroborated by the questionnaire analysis, revealing that the users of \textit{high} cluster w.r.t. MORS ratio provided more positive answers on the questions \textit{``Recommended items were novel to me''}  (significant for all \textit{head} sizes), \textit{``Recommended items were diverse''} (significant for \textit{head} sizes of three and four), and \textit{``Recommended items matched my interests''} (significant for \textit{low-selection} users and \textit{head} sizes of three and four).

\vspace{1mm}
\noindent {\bf Impact of MORS selections on the fine-tuning of single-objective RS.}
In this analysis, we aimed on discovering to what extent was it beneficial to fine-tune single-objective RS with the help of MORS-based selections. To do so, we simulated the behavior of \textit{Beta}, should it be trained only w.r.t. selections made on single-objective recommendations.\footnote{The procedure was incremental, i.e., in each session, we only considered those selections, for which the re-trained \textit{Beta} provided an impression.} First, note that while the recommendations of single-objective-trained \textit{Beta} (\textit{SOT-Beta}) gradually departed from the original \textit{Beta}, the intersection remained substantial (decreasing from 80\% in the second session to 63\% in the last session). This makes the whole procedure feasible, although we can expect that due to the lower volume of impressions, obtained results could somewhat underestimate the true performance of \textit{SOT-Beta}. 

\begin{figure*}
    \centering
    \includegraphics[width=\linewidth]{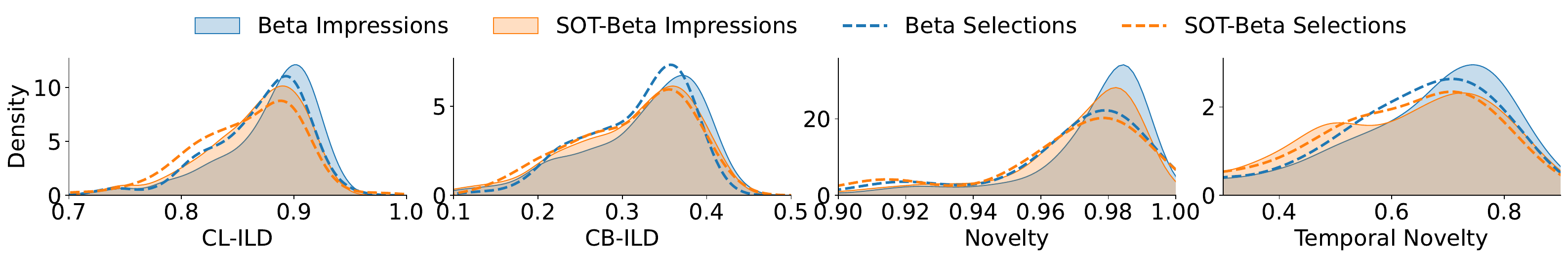}
    \caption{Comparison of Single-objective trained Beta (\textit{SOT-Beta}) and default Beta, w.r.t. collaborative diversity (\textit{CL-ILD}), content-based diversity (\textit{CB-ILD}), novelty, and temporal novelty on impressions and selections.}
    \label{fig:single_objective_beta_properties}
\end{figure*}

Now, let us observe the beyond-accuracy statistics of both \textit{Beta} variants (see Figure \ref{fig:single_objective_beta_properties}). \textit{SOT-Beta} exhibited lower collaborative ILD (0.864 vs 0.876; T-test p-value:3.7e-8) w.r.t. impressions. More importantly, this also translated into the inferior ILD w.r.t. selections -- that is, when comparing the ILD of all selections of original \textit{Beta} recommendations with those recommended \textit{SOT-Beta} (0.857 vs 0.869, p-value: 0.028). Similar observations can be made also for impression-based content-based ILD, genre coverage, mean popularity complement, and temporal novelty. 
In these cases, selection-level statistics were also slightly better for \textit{Beta}, but the difference was not significant. 
To conclude, the existence of MORS-based selections considerably improved the beyond-accuracy properties of the single-objective RS (as long as impressed items are considered), and this partially translated into the improved adoption of items with higher beyond-accuracy statistics by study participants.



\subsection{Comparing user's selections with user-defined propensities towards beyond-accuracy objectives}
As it can be observed from Table \ref{tab:overall_results}, users' selections did not exactly follow the distribution of the impressions. Selections made on \textit{Beta} recommender exhibited significantly lower $NMG_{rel}$ than corresponding impressions (T-test p-value: 4.4e-5). Also, selections made on both MORS exhibited significantly lower $NMG_{div}$ and $NMG_{nov}$ and simultaneously higher $NMG_{rel}$ (p-values < 2.6e-14). Note that while the decrease of selection's $NMG_{rel}$ w.r.t. \textit{Beta} may seem modest, it is due to a very narrow distribution of $NMG_{rel}$ on impressions.

\begin{figure*}
    \centering
    \includegraphics[width=\linewidth]{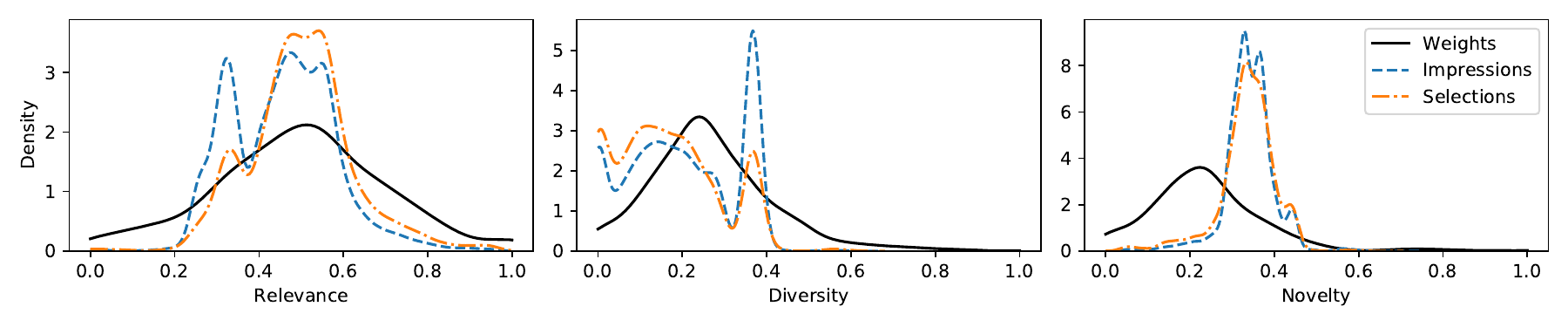}
    \caption{Comparison of distributions for user-defined propensity weights, and relative marginal gains on impressions and selections. }
    \label{fig:weights_impressions_selections_distrib}
\end{figure*}

Differences between selection and impression distributions naturally lead to the question of to what extent this is connected to the users' self-proclaimed propensity weights. Figure \ref{fig:weights_impressions_selections_distrib} depicts distributions of propensity weights and $NMG$s\footnote{In order to make NMGs comparable with propensity weights, we re-scaled them to maintain unit sum object-wise.}  of impressions and selections. Notably, for impressions, the distribution lacks the segment of very high/low values despite the demand expressed in propensity weights. This is mostly due to the existing covariance among selected objectives, which, e.g., prevents from finding diverse items without certain levels of novelty. More importantly, note that both relevance and diversity metrics were under-represented in impressions, if compared to the propensity weights. However, while for relevance, users tend to balance this bias back by only rarely selecting items with very low relevance (see the spike towards the left side of Figure \ref{fig:weights_impressions_selections_distrib}), for diversity, items with low $NMG_{div}$ are over-sampled and thus the difference between selection behavior and self-proclaimed propensity amplifies. 

Despite the fact that users had the freedom of choosing objective weights at their discretion, their selection behavior significantly differs from the self-proclaimed propensity towards diversity, amplifying the existing bias of impression data. Seemingly, users overstated their propensity towards diversity. To some extent, we corroborated this hypothesis by analyzing the satisfaction of users as a factor of their propensity towards diversity. Users with a below-median propensity towards diversity on average selected more items from MORS recommendations (3.28 vs. 2.51, p-value 2.2e-7) and also provided higher overall ratings for MORS recommendations (2.84 vs. 2.64, p-value: 0.001). Originally, we expected that based on these inferior results, users would tend to converge towards lower weights for the diversity objective. However, no such evidence was found in the dataset.

This observation could have several causes. The limited number of sessions might be simply insufficient to learn the dependencies between objective weights and self-perceived satisfaction, let alone that the dependence might vary through time as illustrated in Section \ref{sec:results_long_term}. Nonetheless, the misconception or misunderstanding on the level of objective semantics and/or item's marginal gains w.r.t. these objectives may play an important role too. The post-study questionnaire provided some leads on this factor. User's overall user satisfaction (i.e., \textit{``Overall, I am satisfied with the recommender.''}) was correlated with the information sufficiency (\textit{``The information provided for the recommended movies was sufficient to judge whether I gonna like them.''}, Pearson's correlation: 0.42) and the ability to state one's preferences (\textit{``I was not able to describe my preferences w.r.t. relevance, diversity, and novelty.''}, Pearson's correlation: -0.43). Also, while evaluating the user-perceived fulfillment of individual objectives, we found that positive answers on \textit{``The movies recommended to me matched my interests.''} implied no significant relations to the estimated relevances. Furthermore, while the positive answers on \textit{``The recommended movies were novel to me.''} implied some increase of the novelty metrics (0.985 vs. 0.981 for mean popularity complement and 0.798 vs. 0.751 for temporal novelty), the magnitude  of improvement was much higher for users who answered positively on \textit{``The recommended movies were diverse.''}: 0.986 vs. 0.976 for mean popularity complement and 0.795 vs. 0.716 for temporal novelty. 

We can conclude the level of misconception between objective metrics and users' perception of these qualities is substantial. This is in line with the observations in related studies, e.g., \cite{7994718}. Some parts of the post-study questionnaire suggest that this issue may be mitigated by better explanations (i.e., more informative descriptions) of recommended items. One option would be to visualize the degree, to which items fulfill individual objectives. This would allow users to better link their perception with underlying metrics and, e.g., help to adapt the self-proclaimed propensities to this knowledge.

\section{Conclusions and Limitations}\label{sec:conclusions}
In this paper, we conducted a user study focused on discovering the dependencies between users' interactions on items with certain beyond-accuracy properties and users' self-proclaimed propensities towards these beyond-accuracy criteria. We observed a considerable drift between both statistics and investigated the possible causes. We also provided some evidence of the benefits of MORS, despite not being the favored option from the user's (short-term) perspective.

The study had several limitations, which we plan to address in the future. 
First, only a modest volume of recommending sessions was conducted with no time in between. This prevents us from measuring the preference drifts \cite{9617631} and/or contextual dependencies in long-term impact analysis. 
Also, users might not have enough time to stabilize their propensities towards individual objectives. Therefore, our future work should include studies with longer trial periods and sufficient time in between.  
Second, the choice of beyond-accuracy objectives as well as their particular definitions might affect the results. We plan to address this by a future study with a wider set of beyond-accuracy objectives. Similarly, we plan to investigate the impact of particular UIs for setting propensity weights. 
Third, the fact that \textit{Beta} RS was trained w.r.t. all selections correspond to the situation, where an ensemble model is used. While this is plausible, we would also like to observe the effect of independent evolution for all RS. 

Last but not least, the study was rather limited in terms of its size. Only one dataset, only one relevance-based RS, and only around 100 participants were employed. This might impact the stability of the results. Also, a rather strict filter on train set users and items was employed, resulting in a modest volume of candidate items. As such, it may prove difficult to find suitable recommendations in sequential settings without repetition. To mitigate these limitations, we plan to conduct an extensive set of follow-up studies comprising larger pools of participants, more evaluated RS variants, and more domains, including large-scale ones. 

\begin{acks}
This paper has been supported by Czech Science Foundation (GA\v{C}R) project 22-21696S, Charles University grant SVV-260698/2023, and Charles University Grant Agency (GA UK) project number 188322. We acknowledge financial support under the National Recovery and Resilience Plan (NRRP), Mission 4 Component 2 Investment 1.5 - Call for tender No.3277 published on December 30, 2021 by the Italian Ministry of University and Research (MUR) funded by the European Union – NextGenerationEU. Project Code ECS0000038 – Project Title eINS Ecosystem of Innovation for Next Generation Sardinia – CUP F53C22000430001- Grant Assignment Decree No. 1056 adopted on June 23, 2022 by the Italian Ministry of University and Research (MUR).
\end{acks}

\bibliographystyle{ACM-Reference-Format}
\bibliography{sample-base}


\begin{thebibliography}{23}


\ifx \showCODEN    \undefined \def \showCODEN     #1{\unskip}     \fi
\ifx \showDOI      \undefined \def \showDOI       #1{#1}\fi
\ifx \showISBNx    \undefined \def \showISBNx     #1{\unskip}     \fi
\ifx \showISBNxiii \undefined \def \showISBNxiii  #1{\unskip}     \fi
\ifx \showISSN     \undefined \def \showISSN      #1{\unskip}     \fi
\ifx \showLCCN     \undefined \def \showLCCN      #1{\unskip}     \fi
\ifx \shownote     \undefined \def \shownote      #1{#1}          \fi
\ifx \showarticletitle \undefined \def \showarticletitle #1{#1}   \fi
\ifx \showURL      \undefined \def \showURL       {\relax}        \fi
\providecommand\bibfield[2]{#2}
\providecommand\bibinfo[2]{#2}
\providecommand\natexlab[1]{#1}
\providecommand\showeprint[2][]{arXiv:#2}

\bibitem[Boratto and Marras(2021)]%
        {BorattoM21}
\bibfield{author}{\bibinfo{person}{Ludovico Boratto} {and}
  \bibinfo{person}{Mirko Marras}.} \bibinfo{year}{2021}\natexlab{}.
\newblock \showarticletitle{Advances in Bias-aware Recommendation on the Web}.
  In \bibinfo{booktitle}{\emph{{WSDM} '21, The Fourteenth {ACM} International
  Conference on Web Search and Data Mining, Virtual Event, Israel, March 8-12,
  2021}}, \bibfield{editor}{\bibinfo{person}{Liane Lewin{-}Eytan},
  \bibinfo{person}{David Carmel}, \bibinfo{person}{Elad Yom{-}Tov},
  \bibinfo{person}{Eugene Agichtein}, {and} \bibinfo{person}{Evgeniy
  Gabrilovich}} (Eds.). \bibinfo{publisher}{{ACM}},
  \bibinfo{pages}{1147--1149}.
\newblock
\urldef\tempurl%
\url{https://doi.org/10.1145/3437963.3441665}
\showDOI{\tempurl}


\bibitem[Bradley and Smyth(2001)]%
        {bradley2001improving}
\bibfield{author}{\bibinfo{person}{Keith Bradley} {and} \bibinfo{person}{Barry
  Smyth}.} \bibinfo{year}{2001}\natexlab{}.
\newblock \showarticletitle{Improving recommendation diversity}. In
  \bibinfo{booktitle}{\emph{Proceedings of the twelfth Irish conference on
  artificial intelligence and cognitive science, Maynooth, Ireland}},
  Vol.~\bibinfo{volume}{85}. Citeseer, \bibinfo{pages}{141--152}.
\newblock


\bibitem[Castells et~al\mbox{.}(2022)]%
        {CastellsHV22}
\bibfield{author}{\bibinfo{person}{Pablo Castells}, \bibinfo{person}{Neil
  Hurley}, {and} \bibinfo{person}{Sa{\'{u}}l Vargas}.}
  \bibinfo{year}{2022}\natexlab{}.
\newblock \showarticletitle{Novelty and Diversity in Recommender Systems}.
\newblock In \bibinfo{booktitle}{\emph{Recommender Systems Handbook}},
  \bibfield{editor}{\bibinfo{person}{Francesco Ricci}, \bibinfo{person}{Lior
  Rokach}, {and} \bibinfo{person}{Bracha Shapira}} (Eds.).
  \bibinfo{publisher}{Springer {US}}, \bibinfo{pages}{603--646}.
\newblock
\urldef\tempurl%
\url{https://doi.org/10.1007/978-1-0716-2197-4\_16}
\showDOI{\tempurl}


\bibitem[Elahi et~al\mbox{.}(2014)]%
        {elahiinteractive}
\bibfield{author}{\bibinfo{person}{Mehdi Elahi}, \bibinfo{person}{Mouzhi Ge},
  \bibinfo{person}{Francesco Ricci}, \bibinfo{person}{David Massimo}, {and}
  \bibinfo{person}{Shlomo Berkovsky}.} \bibinfo{year}{2014}\natexlab{}.
\newblock \showarticletitle{Interactive Food Recommendation for Groups}. In
  \bibinfo{booktitle}{\emph{8th ACM Conference on Recommender Systems, RecSys
  2014}}. \bibinfo{publisher}{CEUR-WS}.
\newblock


\bibitem[Fazeli et~al\mbox{.}(2018)]%
        {7994718}
\bibfield{author}{\bibinfo{person}{Soude Fazeli}, \bibinfo{person}{Hendrik
  Drachsler}, \bibinfo{person}{Marlies Bitter-Rijpkema},
  \bibinfo{person}{Francis Brouns}, \bibinfo{person}{Wim van~der Vegt}, {and}
  \bibinfo{person}{Peter~B. Sloep}.} \bibinfo{year}{2018}\natexlab{}.
\newblock \showarticletitle{User-Centric Evaluation of Recommender Systems in
  Social Learning Platforms: Accuracy is Just the Tip of the Iceberg}.
\newblock \bibinfo{journal}{\emph{IEEE Transactions on Learning Technologies}}
  \bibinfo{volume}{11}, \bibinfo{number}{3} (\bibinfo{year}{2018}),
  \bibinfo{pages}{294--306}.
\newblock
\urldef\tempurl%
\url{https://doi.org/10.1109/TLT.2017.2732349}
\showDOI{\tempurl}


\bibitem[Gao et~al\mbox{.}(2021)]%
        {GAO2021100}
\bibfield{author}{\bibinfo{person}{Chongming Gao}, \bibinfo{person}{Wenqiang
  Lei}, \bibinfo{person}{Xiangnan He}, \bibinfo{person}{Maarten {de Rijke}},
  {and} \bibinfo{person}{Tat-Seng Chua}.} \bibinfo{year}{2021}\natexlab{}.
\newblock \showarticletitle{Advances and challenges in conversational
  recommender systems: A survey}.
\newblock \bibinfo{journal}{\emph{AI Open}}  \bibinfo{volume}{2}
  (\bibinfo{year}{2021}), \bibinfo{pages}{100--126}.
\newblock
\showISSN{2666-6510}
\urldef\tempurl%
\url{https://doi.org/10.1016/j.aiopen.2021.06.002}
\showDOI{\tempurl}


\bibitem[Ge et~al\mbox{.}(2022)]%
        {DBLP:conf/wsdm/GeZYPHHZ22}
\bibfield{author}{\bibinfo{person}{Yingqiang Ge}, \bibinfo{person}{Xiaoting
  Zhao}, \bibinfo{person}{Lucia Yu}, \bibinfo{person}{Saurabh Paul},
  \bibinfo{person}{Diane Hu}, \bibinfo{person}{Chu{-}Cheng Hsieh}, {and}
  \bibinfo{person}{Yongfeng Zhang}.} \bibinfo{year}{2022}\natexlab{}.
\newblock \showarticletitle{Toward Pareto Efficient Fairness-Utility Trade-off
  in Recommendation through Reinforcement Learning}. In
  \bibinfo{booktitle}{\emph{{WSDM} '22: The Fifteenth {ACM} International
  Conference on Web Search and Data Mining, Virtual Event / Tempe, AZ, USA,
  February 21 - 25, 2022}}, \bibfield{editor}{\bibinfo{person}{K.~Selcuk
  Candan}, \bibinfo{person}{Huan Liu}, \bibinfo{person}{Leman Akoglu},
  \bibinfo{person}{Xin~Luna Dong}, {and} \bibinfo{person}{Jiliang Tang}}
  (Eds.). \bibinfo{publisher}{{ACM}}, \bibinfo{pages}{316--324}.
\newblock
\urldef\tempurl%
\url{https://doi.org/10.1145/3488560.3498487}
\showDOI{\tempurl}


\bibitem[Harper and Konstan(2015)]%
        {10.1145/2827872}
\bibfield{author}{\bibinfo{person}{F.~Maxwell Harper} {and}
  \bibinfo{person}{Joseph~A. Konstan}.} \bibinfo{year}{2015}\natexlab{}.
\newblock \showarticletitle{The MovieLens Datasets: History and Context}.
\newblock \bibinfo{journal}{\emph{ACM Trans. Interact. Intell. Syst.}}
  \bibinfo{volume}{5}, \bibinfo{number}{4}, Article \bibinfo{articleno}{19}
  (\bibinfo{date}{dec} \bibinfo{year}{2015}), \bibinfo{numpages}{19}~pages.
\newblock
\showISSN{2160-6455}
\urldef\tempurl%
\url{https://doi.org/10.1145/2827872}
\showDOI{\tempurl}


\bibitem[Hu et~al\mbox{.}(2008)]%
        {4781121}
\bibfield{author}{\bibinfo{person}{Yifan Hu}, \bibinfo{person}{Yehuda Koren},
  {and} \bibinfo{person}{Chris Volinsky}.} \bibinfo{year}{2008}\natexlab{}.
\newblock \showarticletitle{Collaborative Filtering for Implicit Feedback
  Datasets}. In \bibinfo{booktitle}{\emph{2008 Eighth IEEE International
  Conference on Data Mining}}. \bibinfo{pages}{263--272}.
\newblock
\urldef\tempurl%
\url{https://doi.org/10.1109/ICDM.2008.22}
\showDOI{\tempurl}


\bibitem[Jannach(2022)]%
        {Jannach22}
\bibfield{author}{\bibinfo{person}{Dietmar Jannach}.}
  \bibinfo{year}{2022}\natexlab{}.
\newblock \showarticletitle{Multi-Objective Recommendation: Overview and
  Challenges}. In \bibinfo{booktitle}{\emph{Proceedings of the 2nd Workshop on
  Multi-Objective Recommender Systems co-located with 16th {ACM} Conference on
  Recommender Systems (RecSys 2022), Seattle, WA, USA, 18th-23rd September
  2022}} \emph{(\bibinfo{series}{{CEUR} Workshop Proceedings},
  Vol.~\bibinfo{volume}{3268})}, \bibfield{editor}{\bibinfo{person}{Himan
  Abdollahpouri}, \bibinfo{person}{Shaghayegh Sahebi}, \bibinfo{person}{Mehdi
  Elahi}, \bibinfo{person}{Masoud Mansoury}, \bibinfo{person}{Babak Loni},
  \bibinfo{person}{Zahra Nazari}, {and} \bibinfo{person}{Maria Dimakopoulou}}
  (Eds.). \bibinfo{publisher}{CEUR-WS.org}.
\newblock
\urldef\tempurl%
\url{https://ceur-ws.org/Vol-3268/paper1.pdf}
\showURL{%
\tempurl}


\bibitem[Jugovac et~al\mbox{.}(2017)]%
        {JUGOVAC2017321}
\bibfield{author}{\bibinfo{person}{Michael Jugovac}, \bibinfo{person}{Dietmar
  Jannach}, {and} \bibinfo{person}{Lukas Lerche}.}
  \bibinfo{year}{2017}\natexlab{}.
\newblock \showarticletitle{Efficient optimization of multiple recommendation
  quality factors according to individual user tendencies}.
\newblock \bibinfo{journal}{\emph{Expert Systems with Applications}}
  \bibinfo{volume}{81} (\bibinfo{year}{2017}), \bibinfo{pages}{321--331}.
\newblock
\showISSN{0957-4174}
\urldef\tempurl%
\url{https://doi.org/10.1016/j.eswa.2017.03.055}
\showDOI{\tempurl}


\bibitem[Li et~al\mbox{.}(2018)]%
        {10.5555/3327546.3327641}
\bibfield{author}{\bibinfo{person}{Raymond Li}, \bibinfo{person}{Samira Kahou},
  \bibinfo{person}{Hannes Schulz}, \bibinfo{person}{Vincent Michalski},
  \bibinfo{person}{Laurent Charlin}, {and} \bibinfo{person}{Chris Pal}.}
  \bibinfo{year}{2018}\natexlab{}.
\newblock \showarticletitle{Towards Deep Conversational Recommendations}. In
  \bibinfo{booktitle}{\emph{Proceedings of the 32nd International Conference on
  Neural Information Processing Systems}} (Montr\'{e}al, Canada)
  \emph{(\bibinfo{series}{NIPS'18})}. \bibinfo{publisher}{Curran Associates
  Inc.}, \bibinfo{address}{Red Hook, NY, USA}, \bibinfo{pages}{9748–9758}.
\newblock


\bibitem[Li et~al\mbox{.}(2021)]%
        {DBLP:conf/www/LiCFGZ21}
\bibfield{author}{\bibinfo{person}{Yunqi Li}, \bibinfo{person}{Hanxiong Chen},
  \bibinfo{person}{Zuohui Fu}, \bibinfo{person}{Yingqiang Ge}, {and}
  \bibinfo{person}{Yongfeng Zhang}.} \bibinfo{year}{2021}\natexlab{}.
\newblock \showarticletitle{User-oriented Fairness in Recommendation}. In
  \bibinfo{booktitle}{\emph{{WWW} '21: The Web Conference 2021, Virtual Event /
  Ljubljana, Slovenia, April 19-23, 2021}},
  \bibfield{editor}{\bibinfo{person}{Jure Leskovec}, \bibinfo{person}{Marko
  Grobelnik}, \bibinfo{person}{Marc Najork}, \bibinfo{person}{Jie Tang}, {and}
  \bibinfo{person}{Leila Zia}} (Eds.). \bibinfo{publisher}{{ACM} / {IW3C2}},
  \bibinfo{pages}{624--632}.
\newblock
\urldef\tempurl%
\url{https://doi.org/10.1145/3442381.3449866}
\showDOI{\tempurl}


\bibitem[Lin et~al\mbox{.}(2019)]%
        {DBLP:conf/recsys/LinCPSXSZOJ19}
\bibfield{author}{\bibinfo{person}{Xiao Lin}, \bibinfo{person}{Hongjie Chen},
  \bibinfo{person}{Changhua Pei}, \bibinfo{person}{Fei Sun},
  \bibinfo{person}{Xuanji Xiao}, \bibinfo{person}{Hanxiao Sun},
  \bibinfo{person}{Yongfeng Zhang}, \bibinfo{person}{Wenwu Ou}, {and}
  \bibinfo{person}{Peng Jiang}.} \bibinfo{year}{2019}\natexlab{}.
\newblock \showarticletitle{A pareto-efficient algorithm for multiple objective
  optimization in e-commerce recommendation}. In
  \bibinfo{booktitle}{\emph{Proceedings of the 13th {ACM} Conference on
  Recommender Systems, RecSys 2019, Copenhagen, Denmark, September 16-20,
  2019}}, \bibfield{editor}{\bibinfo{person}{Toine Bogers},
  \bibinfo{person}{Alan Said}, \bibinfo{person}{Peter Brusilovsky}, {and}
  \bibinfo{person}{Domonkos Tikk}} (Eds.). \bibinfo{publisher}{{ACM}},
  \bibinfo{pages}{20--28}.
\newblock
\urldef\tempurl%
\url{https://doi.org/10.1145/3298689.3346998}
\showDOI{\tempurl}


\bibitem[Naghiaei et~al\mbox{.}(2022)]%
        {DBLP:conf/sigir/NaghiaeiRD22}
\bibfield{author}{\bibinfo{person}{Mohammadmehdi Naghiaei},
  \bibinfo{person}{Hossein~A. Rahmani}, {and} \bibinfo{person}{Yashar
  Deldjoo}.} \bibinfo{year}{2022}\natexlab{}.
\newblock \showarticletitle{CPFair: Personalized Consumer and Producer Fairness
  Re-ranking for Recommender Systems}. In \bibinfo{booktitle}{\emph{{SIGIR}
  '22: The 45th International {ACM} {SIGIR} Conference on Research and
  Development in Information Retrieval, Madrid, Spain, July 11 - 15, 2022}},
  \bibfield{editor}{\bibinfo{person}{Enrique Amig{\'{o}}},
  \bibinfo{person}{Pablo Castells}, \bibinfo{person}{Julio Gonzalo},
  \bibinfo{person}{Ben Carterette}, \bibinfo{person}{J.~Shane Culpepper}, {and}
  \bibinfo{person}{Gabriella Kazai}} (Eds.). \bibinfo{publisher}{{ACM}},
  \bibinfo{pages}{770--779}.
\newblock
\urldef\tempurl%
\url{https://doi.org/10.1145/3477495.3531959}
\showDOI{\tempurl}


\bibitem[Peska and Dokoupil(2022)]%
        {10.1145/3477495.3531787}
\bibfield{author}{\bibinfo{person}{Ladislav Peska} {and}
  \bibinfo{person}{Patrik Dokoupil}.} \bibinfo{year}{2022}\natexlab{}.
\newblock \showarticletitle{Towards Results-Level Proportionality for
  Multi-Objective Recommender Systems}. In
  \bibinfo{booktitle}{\emph{Proceedings of the 45th International ACM SIGIR
  Conference on Research and Development in Information Retrieval}} (Madrid,
  Spain) \emph{(\bibinfo{series}{SIGIR '22})}. \bibinfo{publisher}{Association
  for Computing Machinery}, \bibinfo{address}{New York, NY, USA},
  \bibinfo{pages}{1963–1968}.
\newblock
\showISBNx{9781450387323}
\urldef\tempurl%
\url{https://doi.org/10.1145/3477495.3531787}
\showDOI{\tempurl}


\bibitem[Pu et~al\mbox{.}(2011)]%
        {10.1145/2043932.2043962}
\bibfield{author}{\bibinfo{person}{Pearl Pu}, \bibinfo{person}{Li Chen}, {and}
  \bibinfo{person}{Rong Hu}.} \bibinfo{year}{2011}\natexlab{}.
\newblock \showarticletitle{A User-Centric Evaluation Framework for Recommender
  Systems}. In \bibinfo{booktitle}{\emph{Proceedings of the Fifth ACM
  Conference on Recommender Systems}} (Chicago, Illinois, USA)
  \emph{(\bibinfo{series}{RecSys '11})}. \bibinfo{publisher}{Association for
  Computing Machinery}, \bibinfo{address}{New York, NY, USA},
  \bibinfo{pages}{157–164}.
\newblock
\showISBNx{9781450306836}
\urldef\tempurl%
\url{https://doi.org/10.1145/2043932.2043962}
\showDOI{\tempurl}


\bibitem[Sritrakool and Maneeroj(2021)]%
        {9617631}
\bibfield{author}{\bibinfo{person}{Nakarin Sritrakool} {and}
  \bibinfo{person}{Saranya Maneeroj}.} \bibinfo{year}{2021}\natexlab{}.
\newblock \showarticletitle{Personalized Preference Drift Aware Sequential
  Recommender System}.
\newblock \bibinfo{journal}{\emph{IEEE Access}}  \bibinfo{volume}{9}
  (\bibinfo{year}{2021}), \bibinfo{pages}{155491--155506}.
\newblock
\urldef\tempurl%
\url{https://doi.org/10.1109/ACCESS.2021.3128769}
\showDOI{\tempurl}


\bibitem[Stamenkovic et~al\mbox{.}(2022)]%
        {DBLP:conf/wsdm/StamenkovicKAXK22}
\bibfield{author}{\bibinfo{person}{Dusan Stamenkovic},
  \bibinfo{person}{Alexandros Karatzoglou}, \bibinfo{person}{Ioannis Arapakis},
  \bibinfo{person}{Xin Xin}, {and} \bibinfo{person}{Kleomenis Katevas}.}
  \bibinfo{year}{2022}\natexlab{}.
\newblock \showarticletitle{Choosing the Best of Both Worlds: Diverse and Novel
  Recommendations through Multi-Objective Reinforcement Learning}. In
  \bibinfo{booktitle}{\emph{{WSDM} '22: The Fifteenth {ACM} International
  Conference on Web Search and Data Mining, Virtual Event / Tempe, AZ, USA,
  February 21 - 25, 2022}}, \bibfield{editor}{\bibinfo{person}{K.~Selcuk
  Candan}, \bibinfo{person}{Huan Liu}, \bibinfo{person}{Leman Akoglu},
  \bibinfo{person}{Xin~Luna Dong}, {and} \bibinfo{person}{Jiliang Tang}}
  (Eds.). \bibinfo{publisher}{{ACM}}, \bibinfo{pages}{957--965}.
\newblock
\urldef\tempurl%
\url{https://doi.org/10.1145/3488560.3498471}
\showDOI{\tempurl}


\bibitem[Vargas and Castells(2011)]%
        {epc}
\bibfield{author}{\bibinfo{person}{Sa\'{u}l Vargas} {and}
  \bibinfo{person}{Pablo Castells}.} \bibinfo{year}{2011}\natexlab{}.
\newblock \showarticletitle{Rank and Relevance in Novelty and Diversity Metrics
  for Recommender Systems}. In \bibinfo{booktitle}{\emph{Proceedings of the
  Fifth ACM Conference on Recommender Systems}} (Chicago, Illinois, USA)
  \emph{(\bibinfo{series}{RecSys '11})}. \bibinfo{publisher}{Association for
  Computing Machinery}, \bibinfo{address}{New York, NY, USA},
  \bibinfo{pages}{109–116}.
\newblock
\showISBNx{9781450306836}
\urldef\tempurl%
\url{https://doi.org/10.1145/2043932.2043955}
\showDOI{\tempurl}


\bibitem[Wang et~al\mbox{.}(2020)]%
        {WANG2020106369}
\bibfield{author}{\bibinfo{person}{Zhaoyuan Wang}, \bibinfo{person}{Chuishi
  Meng}, \bibinfo{person}{Shenggong Ji}, \bibinfo{person}{Tianrui Li}, {and}
  \bibinfo{person}{Yu Zheng}.} \bibinfo{year}{2020}\natexlab{}.
\newblock \showarticletitle{Food package suggestion system based on
  multi-objective optimization: A case study on a real-world restaurant}.
\newblock \bibinfo{journal}{\emph{Applied Soft Computing}}
  \bibinfo{volume}{93} (\bibinfo{year}{2020}), \bibinfo{pages}{106369}.
\newblock
\showISSN{1568-4946}
\urldef\tempurl%
\url{https://doi.org/10.1016/j.asoc.2020.106369}
\showDOI{\tempurl}


\bibitem[Wu et~al\mbox{.}(2022)]%
        {wu2022multifr}
\bibfield{author}{\bibinfo{person}{Haolun Wu}, \bibinfo{person}{Chen Ma},
  \bibinfo{person}{Bhaskar Mitra}, \bibinfo{person}{Fernando Diaz}, {and}
  \bibinfo{person}{Xue Liu}.} \bibinfo{year}{2022}\natexlab{}.
\newblock \showarticletitle{Multi-FR: A Multi-objective Optimization Framework
  for Multi-stakeholder Fairness-aware Recommendation}. In
  \bibinfo{booktitle}{\emph{Transactions on Information Systems (TOIS)}}.
  \bibinfo{publisher}{{ACM}}.
\newblock


\bibitem[Zheng and Wang(2022)]%
        {ZhengW22}
\bibfield{author}{\bibinfo{person}{Yong Zheng} {and}
  \bibinfo{person}{David~(Xuejun) Wang}.} \bibinfo{year}{2022}\natexlab{}.
\newblock \showarticletitle{A survey of recommender systems with
  multi-objective optimization}.
\newblock \bibinfo{journal}{\emph{Neurocomputing}}  \bibinfo{volume}{474}
  (\bibinfo{year}{2022}), \bibinfo{pages}{141--153}.
\newblock
\urldef\tempurl%
\url{https://doi.org/10.1016/j.neucom.2021.11.041}
\showDOI{\tempurl}


\end{thebibliography}

\end{document}